\documentclass{article}

\usepackage{arxiv}
\usepackage{graphicx}
\title{Minimum Conversion Sets of $f$-reversible Processes on Paths}
\author{%
  \href{https://orcid.org/0000-0001-9964-8816}{%
    \includegraphics[height=0.8em]{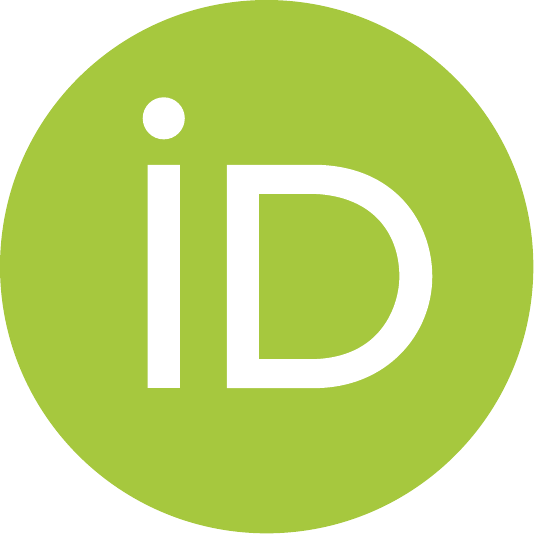}%
    \hspace{1mm}%
    Volker Turau%
  }  \\
  Hamburg University of Technology \\
  21073 Hamburg, Germany \\
  \texttt{turau@tuhh.de} 
      }

\usepackage[utf8]{inputenc}
\usepackage[english]{babel}

\usepackage{amsmath}
\usepackage{amssymb}      
\usepackage{amsthm}
\usepackage{xspace}
\usepackage{mathbbol}
\usepackage[e]{esvect}
\usepackage{hyperref}

\newtheorem{theorem}{Theorem}
\newtheorem{lemma}{Lemma}
\newtheorem{definition}{Definition}
\newtheorem{corollary}{Corollary}

\newcommand{\cev}[1]{\reflectbox{\ensuremath{\vec{\reflectbox{\ensuremath{#1}}}}}}

\newcommand{\Cf}{\ensuremath{{\cal C}_f}\xspace}
\newcommand{\Cfp}[1]{\ensuremath{{\cal C}_f(#1)}\xspace}
\newcommand{\CSf}{\ensuremath{{\cal C}_f^m}\xspace} 

\makeatletter
\DeclareRobustCommand{\cev}[1]{%
  {\mathpalette\do@cev{#1}}%
}
\newcommand{\do@cev}[2]{%
  \vbox{\offinterlineskip
    \sbox\z@{$\m@th#1 x$}%
    \ialign{##\cr
      \hidewidth\reflectbox{$\m@th#1\vec{}\mkern4mu$}\hidewidth\cr
      \noalign{\kern-\ht\z@}
      $\m@th#1#2$\cr
    }%
  }%
}
\makeatother

\RequirePackage{mleftright} 
\RequirePackage{xparse}
\RequirePackage{mathtools}

\newcommand{\suchthat}{\;\ifnum\currentgrouptype=16 \middle\fi|\;}

\NewDocumentCommand{\abs}{som}{%
  \IfNoValueTF{#2}{
    \IfBooleanTF{#1}{%
      \left\lvert\mskip0.3\thinmuskip#3\mskip0.3\thinmuskip\right\rvert%
    }{%
      \lvert\mskip0.3\thinmuskip#3\mskip0.3\thinmuskip\rvert%
    }%
  }{%
    \mathopen{#2\lvert}\mskip0.3\thinmuskip#3\mskip0.3\thinmuskip\mathclose{#2\rvert}
  }%
}%

\begin{document}
\maketitle      
%

%
%
\begin{abstract}
  We study $f$-reversible processes on a path $P_n$, in which
  synchronously each node $v$ flips its label whenever at least $f(v)$
  of its neighbors disagree with it, for an arbitrary threshold
  function $f$. The central parameter is the minimum $f$-conversion
  set number $r_f(P_n)$, the smallest initial seed set guaranteeing
  eventual convergence to full activation, i.e., all nodes reach label
  $1$. We prove the lower bound
  $r_f(P_n) \ge \left\lceil n/2\right\rceil$, valid for every choice
  of $f$, and completely classify threshold functions attaining each
  extreme: those for which $r_f(P_n) = \left\lceil n/2\right\rceil$
  and those for which the process degenerates to $r_f(P_n) = n$. These
  results give an exact, non-asymptotic picture of extremal conversion
  behavior on the sparsest connected graphs, complementing existing
  hardness and constant-threshold results in the literature. Finally,
  we prove that the number of threshold functions $f$ with
  $r_f(P_n) = n$ can be computed in time $O(n)$ and that these
  threshold functions can be enumerated in time $O(n + I)$, where $I$
  is the number of such threshold functions.
\end{abstract}

\section{Introduction}
\label{sec:intro}
In this work we consider synchronous dynamical systems $(G,f)$, where
$G = (V, E)$ is a finite graph and $f : V \longrightarrow \mathbb{N}$
a \emph{threshold function}. Starting from an initial labeling
$c: V \longrightarrow \{0, 1\}$, every node $v\in V$ changes its label
in discrete-time steps -- called \emph{rounds} -- according to a local
rule based on $f$, such that every node $v\in V$ changes its label if
and only if it has at least $f(v)$ neighbors with a different label.
An initial labeling $c^0$ uniquely defines a sequence of
configurations $c^0,c^1,\ldots$, where $c^{t-1}(v)$ denotes the label
of $v$ at the beginning of round $t$:
\[c^t(v) = 
\begin{cases}
1 -c^{t-1}(v), &\text{    if } \abs{\{u\in N(v) \suchthat c^{t-1}(v)\not=
c^{t-1}(u)\}}\ge f(v),\\
c^{t-1}(v), &\text{    otherwise.}
\end{cases}\]

This dynamical system $(G,f)$ is called an $f$-\emph{reversible
  process}. Unlike the classical irreversible threshold processes used
to model the spread of disease, opinion, or faults in a network
\cite{Centeno:2011,Baste:2022,Dreyer2009}, a node here is never
permanently committed to a state: it can flip back and forth as the
configuration of its neighborhood changes. This makes $f$-reversible
processes a natural graph-theoretic analogue of consensus and
coordination dynamics in which agents continue to revise their state
in response to their neighbors -- closer in spirit to voter-models and
SIS-type dynamics, where individuals move from {\em Susceptible} to
{\em Infectious} and return directly to {\em Susceptible}, than to the
SIR-type, one-way spread captured by irreversible models.
$f$-reversible processes are employed to model processes in a wide
variety of fields such as social influence \cite{Ning2009}, gene
expression networks \cite{Huang1999}, opinion and disease
dissemination \cite{Dreyer2009}, and distributed computing
\cite{Flocchini2003,Peleg2002}. The local majority process with
$f(v)=\left\lfloor \deg(v)/2\right\rfloor +1$ is an example for a
reversible process \cite{Zehmakan:2019,Turau:2024}.

It is well known that for every initial configuration $c^0$ each
$f$-reversible process reaches a configuration, which is either a
fixed point or a 2-cycle \cite{Goles:1980,Poljak:1983}. In the first
case, we say that $c^0$ \emph{converges} for $f$ to the fixed point,
or in short converges.

For a configuration $c$ of $G$ denote by $\gamma(c)$ the number of
nodes $v$ with $c(v)=1$. For a subset $U$ of $V$, the configuration
$c_U$ is defined as $c_U(v) = 1$ if and only if $v\in U$, so
$\gamma(c_U)=\abs{U}$. Let $\mathbb{1}_G =c_V$. A subset
$U \subseteq V$ is called a \emph{conversion set} of $(G,f),$ if the
configuration $c_U$ converges for $f$ to $\mathbb{1}_G$. Such sets are
also known as dynamic monopolies, or dynamos \cite{Dourado:2012}. $V$
is called the \emph{trivial} conversion set. The central quantity of
interest is the {\em conversion number}
$$ r_f(G) = \min\{\abs{S}\suchthat S \text{ is an } \text{conversion set of }
(G,f)\}$$ written $r_k(G)$ when $f \equiv k$ is constant. Clearly,
$r_f(G) \le n$. Determining $r_f(G)$ for general $G$ and $f$ is
computationally hard: Dreyer showed that deciding $r_k(G) \le q$ is
NP-complete for $k$-regular graphs with $k \ge 3$ \cite{Dreyer2009},
and this hardness persists even on bipartite graphs of maximum degree
$3$ once thresholds are allowed to vary node by node
\cite{Costa:2026}. On the structural side, when $f(v) = d(v)$ for
every node, $r_f(G)$ is in between the node cover number
$\beta(G)$ and $\beta(G)+1$ \cite{Lima:2018}, one of the few general
bounds relating $r_f$ to a classical graph parameter. A closely
related and even more restrictive notion, the $f$-critical set --
where convergence must occur within a single round -- has only very
recently had its parameterized complexity investigated
\cite{marcilon2025,Silva2025}. Denote by
$r^c_f(G) = \min\{\abs{S}\suchthat S \text{ is an } f-\text{critical
  set of } (G,f)\}$. Clearly, $r_f(G) \le r^c_f(G)$.

Dourado et al.\ proved in 2012 the NP-hardness of determining $r_2(G)$
for $(G,f)$ with $G$ a general graph \cite{Dourado:2012}. In 2018 Lima
et al.\ proved the NP-hardness of computing $r_f(G)$, even for
bipartite graphs of maximum degree $3$ when the thresholds of all
nodes are in $\{1, 2, 3\}$ \cite{Lima:2018}. They also proved that
determining whether $r_f (G) \le k$ is NP-complete, even if $G$ is a
planar cubic graph and all thresholds have value $3$. If $G$ is
connected, bipartite and $f(v) = \deg(v)$ for each node $v$, then
$r_f(G)$ can be determined in polynomial time \cite{Dourado:2015}.
Dourado et al.\ presented an algorithm computing $r_f(G)$ for paths
and cycles provided each node with threshold $1$ has
at least one neighbor with threshold $1$ \cite{Dourado:2012}.

In this paper we consider the conversion set problem of a path $P_n$.
Recently, Costa et al.\ presented an algorithm that computes
$r_f(P_n)$ in $O(n^4)$ time \cite{Costa:2026}. They also showed that the
conversion set problem is W[1]-hard when parameterized by the
treewidth and the size of the solution. Our focus is
mainly on the following questions: What is a lower bound for
$r_f(P_n)$? Which threshold functions of $P_n$ only have the trivial
conversion set and which have the smallest conversion sets?

Our contributions are as follows. We prove that
$r_f(P_n)\ge \left\lceil n/2\right\rceil$ for all threshold functions
$f$ (Theorem~\ref{theo:lowerBound}). We also show that this lower
bound is sharp. Therefore,
$r_f(P_n) \in [\left\lceil n/2\right\rceil,n]$ for each $f$.
Furthermore, we characterize threshold functions $f$ with a conversion
set number at the boundary of this interval, i.e.,
$r_f(P_n)=\left\lceil n/2\right\rceil$ or $r_f(P_n)=n$. First, we
characterize threshold functions that have no non-trivial conversion
set (Theorem~\ref{theo:bound}) and show that their number can be
computed in linear time (Theorem~\ref{theo:time_1}). Next, we give a
detailed characterization of threshold functions $f$ that reside at
the other end of the spectrum, i.e., those achieving the lowest
possible conversion set number (Theorem~\ref{cor:cor1}
and~\ref{theo:n_half_fin}).

These results settle the extremal behavior of $r_f(P_n)$ over all
threshold functions $f$, complementing the constant-threshold and
hardness results available in the literature
\cite{Dreyer2009,Dourado:2012,Lima:2018} with an exact, non-asymptotic
characterization on this fundamental graph class. Because of the
general intractability, exact results are typically obtainable only on
structurally simple graph classes, and paths and cycles have
historically served as the natural starting point. Efficient
algorithms for computing dynamos are known for paths and cycles even
where the general problem is NP-hard \cite{Dourado:2012}, and extremal
or exact values of related conversion numbers have likewise been
worked out first for paths before richer classes
\cite{Dreyer2009,Adams2011}.

There are several reasons $P_n$ deserves attention, beyond serving as
a stepping stone toward harder graphs. Firstly, $P_n$ is the graph
with the least possible connectivity among connected graphs, and it is
exactly this sparsity that makes the behavior of $r_f(P_n)$
transparent enough to characterize completely, rather than merely
bound. A full classification of which threshold functions $f$ force
$r_f(P_n)$ to its maximum value $n$ or collapse it to its minimum
$ \left\lceil n/2\right\rceil$ gives a precise picture of how
threshold assignments trade off against topology to determine how much
``seed'' influence is needed to bring an entire network to consensus.

Secondly, line topologies arise directly in modeling linear
communication networks, bus architectures, one-dimensional cellular
automata, and temporal or positional orderings of agents. In each of
these settings, an $f$-reversible process models a scenario, where a
unit adopts a new state (e.g., a faulty/non-faulty status that can
self-correct) whenever a threshold number of its immediate neighbors
already hold that state, and where recovery is possible. Understanding
exactly how much initial ``activation'' $r_f(P_n)$ is required -- and,
crucially, which threshold profiles are cheapest versus most expensive
to convert -- has direct bearing on questions such as the minimum
number of seed nodes needed to guarantee eventual full activation, or,
dually, on identifying threshold assignments making a linear
network maximally resistant to being converted by a small seed set.

Thirdly, the lower bound for $r_f(P_n)$ is of independent interest: it
shows that, no matter how thresholds are chosen, a path can never be
converted by a vanishingly small seed set, in sharp contrast to graphs
of higher connectivity (e.g.\ complete or dense graphs) where small
dynamos exist even for non-trivial thresholds.

\section{Notation}
\label{sec:notation}
In the following $(P_n,f)$ denotes an $f$-reversible process, here
$P_n$ is a path with nodes $v_1,\ldots,v_n$ and
$f : V \longrightarrow \mathbb{N}$ a threshold function of $P_n$.
Denote by $\Cf(P_n)$ -- or \Cf in short -- the set of configurations
of $P_n$ that converge for $f$ to $\mathbb{1}_G$. Furthermore, let
$\Cfp{i} =\{c\in \Cf \mid \gamma(c)=i\}$ and $\CSf={\Cfp{r_f(P_n)}}$.
Let $c$ be a configuration of $(P_n,f)$. If $c\in \Cf$, then any
configuration succeeding $c$ also converges. To simplify notation, we
omit the superscript identifying the round number. We abbreviate
$f(v_i)$ with $f_i$ and $c(v_i)$ with $c_i$. The inverse of a
threshold function $f$ is the function $f'$ with $f'_i = f_{n+1-i}$
for all $i$. Since we are primarily interested in conversion sets of
$P_n$ we restrict the class of threshold functions. If $f_i> \deg(v_i)$
then the initial label of $v_i$ does not change during the whole
process, thus if $c\in \Cf$ then $c_i=1$. In this case each conversion
set of $(P_n,f)$ is equal to the union of a conversion set of
$(\cev{P}_{{i}},f^1)$ and one of $(\vec{P}_{{i}},f^2)$, where
$f^{1}_i=f^{2}_1=2$ and $f^{1}_j=f_j$, $f^{2}_j=f_{i+j-1}$ for
$j\not= i$. Thus,
$r_f(P) = r_{f^1}(\cev{P}_{{i}}) + r_{f^2}(\vec{P}_{{i}}) - 1$. Here,
$\cev{P}_{{i}}$ (resp.\ $\vec{P}_{{i}}$) denotes the subpath with
nodes $v_1,\ldots,v_i$ (resp.\ $v_i,\ldots,v_n$). Therefore, we assume
in this work $f_i\le 2$ for all $i$. On the other hand $f_i = 0$
implies $\Cf = \emptyset$.
Therefore, we only consider threshold functions $f$ with
$f_i\in \{1,2\}$, a common assumption in the literature, e.g.,
\cite{Costa:2026,Lima:2018}.

We define $n$ threshold functions
$F^{1,n},\ldots,F^{n,n}$ such that $F^{i,n}_j=1$ if $i=j$ and
$F^{i,n}_j=2$ otherwise. Let $F^n$ be the threshold function with
$F^n_i=2$ for all $i=1,\ldots,n$ and $C^i$ the labeling with
$C^i(i)=0$ and $C^i(j)=1$ for $j\not= i$.

Since $P_n$ is a path, we regard $f$ as a sequence
$(f_1,\ldots, f_n)$. This allows to express $f$ as a concatenation
$S \circ T$ of subsequences $S=(f_1,\ldots, f_s)$,
$T=(f_{s+1},\ldots,f_n)$ and to use regular expressions to define
particular threshold functions. We use the following standard
quantifiers: $\ast$ indicates zero or more occurrences, $+$ one or
more, and $?$ one or zero occurrences. If $P'$ is a subpath of $P_n$,
then $f|_{P'}$ (resp.\ $c|_{P'}$) denotes the restriction of $f$
(resp.\ $c$) to $P'$. Clearly, $(P',f|_{P'})$ is an
$f|_{P'}$-reversible process on $P'$, $c|_{p'}$ a configuration of
$P'$, and $\gamma(c|_{P'})=\abs{\{v \in V_{P'}\suchthat c(v)=1\} }$.
For $i\le j$, let $P_{i,j}$ be the subpath of $P_n$ with nodes
$v_i,\ldots v_j$. Let $\gamma_{i,j}(c)= \gamma(c|_{P_{i,j}})$. If
$j<i$ then $\gamma_{i,j}(c)=0$. Note that $c\in \Cf$ does not
necessary imply $c_{P'} \in C_{f|_{P'}}$. To simplify notation we omit
the subscript $|_{P'}$ whenever it can be unambiguously determined
from the context.

\section{A lower Bound for the Conversion Set Number}
\label{sec:basic-properties-f}

In this section we will prove that for each path $P_n$ a conversion
set contains independently of the threshold function $f$ at least
$\left\lceil n/2\right\rceil$ nodes. The
following two simple observations are used throughout this work. Let
$c\in \Cf$.
\begin{itemize}
\item If $f_i=f_{i+1}=1$ then $c_i=c_{i+1}$ in all  rounds.
\item If $f_1=2$ (resp.\ $f_n=2$) then clearly $c_1=1$ (resp.\
  $c_n=1$).
\end{itemize}
First, we introduce an important
concept frequently used in this work: \emph{1-streaks}. Basically, a
1-streak is a non-extensible sequence of nodes with threshold $1$. 
\begin{definition}
  Let $(P_n,f)$ an $f$-reversible process. $f$ has a \emph{1-streak}
  of length $l\ge 1$ and index $i$, if $f_i=\ldots= f_{i+l-1}=1$ and
  either $i=1$ or $f_{i-1}=2$ and either $i+l-1=n$ or $f_{i+l}=2$. A
  1-streak of length $l\ge 1$ is called \emph{saturated} by
  $c\in \Cf$, if $c_{i}=\ldots c_{i+l-1}=1$ and $c_{i-1}=1$ (resp.\
  $c_{i+l}=1$) provided node $v_{i-1}$ (resp.\ $v_{i+l}$) exists. $f$
  has a \emph{1-streak} of length $0$ and index $i$, if
  $f_i=f_{i+1}=2$, it is called \emph{saturated} by $c\in \Cf$ if
  $c_{i}=c_{i+1}=1$.
\end{definition}

The following table shows five threshold functions and for each a
sequence of configuration $c^0,c^1,\ldots$. In the first four cases
$c^0$ converges to $\mathbb{1}$, it does not converge in the last case.
The first threshold function contains a 1-streak of length $0$ and
index $4$ that is saturated by $c^0$. The third and fourth contain a
1-streak of length $2$ that is not saturated by $c^0$.

\begin{center}
$f$:~\hspace*{2mm} 2,1,2,2,2 \hspace*{5mm} 2,2,1,2,2,2 \hspace*{6mm}
1,2,1,1,2 \hspace*{6mm} 2,2,2,1,1,2,2,2\hspace*{6mm} 2,1,1,1,2,1,1 \\[3mm]
$c^0$:\hspace*{2mm} 1,0,0,1,1 \hspace*{5mm} 1,0,0,1,0,1 \hspace*{6mm}
1,1,0,0,1 \hspace*{6mm} 1,0,1,0,0,1,0,1\hspace*{6mm} 1,1,1,1,1,0,0\\
$c^1$:\hspace*{2mm} 1,1,0,1,1 \hspace*{5mm} 1,0,1,0,1,1 \hspace*{6mm} 1,1,1,1,1 \hspace*{6mm} 1,1,0,1,1,0,1,1\hspace*{6mm} 1,1,1,1,1,1,0\\
$c^2$:\hspace*{2mm} 1,0,1,1,1 \hspace*{5mm} 1,1,0,1,1,1 \hspace*{26.9mm} 1,1,1,0,0,1,1,1\hspace*{6mm} 1,1,1,1,1,0,1\\
$c^3$:\hspace*{2mm} 1,1,1,1,1 \hspace*{5mm} 1,1,1,1,1,1
\hspace*{26.9mm} 1,1,1,1,1,1,1,1\hspace*{6mm} 1,1,1,1,1,1,0   
\end{center}

Saturated 1-streaks are a useful tool in induction proofs. Consider a
threshold function $f$ with a 1-streak of length $l\ge 1$ and index
$i$ that is saturated by $c\in \Cf$. The key property is that the
nodes $v_{i-1},\ldots,v_{l+i}$ do not change their label while
converging to $\mathbb{1}$. This yields that ${c|_{\cev{P}_{i-1}}}$
(resp.\ ${c|_{\vec{P}_{l+i}}}$) converge on $\cev{P}_{i-1}$ (resp.
${\vec{P}_{l+i}}$) for ${f|_{\cev{P}_{i-1}}}$ (resp.\
${f|_{\vec{P}_{l+i}}}$) to $\mathbb{1}$. For 1-streaks of length $0$
the same statement holds for $\cev{P}_{i}$ and ${\vec{P}_{i+1}}$.
Lemma~\ref{lem:splittable} and \ref{lem:splittable_2} leverage this
property of 1-streaks.

\begin{lemma}[Appendix~\ref{sec:proof-lemma-splittable}]\label{lem:splittable}
  Let $(P_n,f)$ be an $f$-reversible process with a 1-streak of length
  $l$ and index $i$ that is saturated by $c\in \CSf$. If $l\ge 1$ then
  $r_f(P) =r_{f|_{\cev{P}_{i-1}}}(\cev{P}_{i-1}) + l +
  r_{f|_{\vec{P}_{l+i}}}(\vec{P}_{l+i})$. If $l=0$ then
  $r_f(P) =r_{f|_{\cev{P}_{i}}}(\cev{P}_{i}) +
  r_{f|_{\vec{P}_{i+1}}}(\vec{P}_{i+1})$.
\end{lemma}

For unsaturated 1-streaks of length 2 we prove a similar result.
\begin{lemma}[Appendix~\ref{sec:proof-lemma-splittable_2}]\label{lem:splittable_2}
  Let $(P_n,f)$ be an $f$-reversible process. If there exists a
  1-streak of length $2$ and index $i$ then
  $r_f(P_n) \ge r_{f|_{\cev{P}_{i}}}(\cev{P}_{i}) +
  r_{f|_{\vec{P}_{i+1}}}(\vec{P}_{i+1})$.
\end{lemma}

The inequality in the last lemma can be strict. Consider a path $P$ of
length $7$ and $f=(2,2,1,1,2,2,2)$, then $r_f(P)=5$ (use
$c=(1,0,1,1,0,1,1)$), but
$r_{f|_{\cev{P}_{3}}}(\cev{P}_{3}) = r_{f|_{\vec{P}_{4}}}(\vec{P}_{4})
=2$. The next Lemma describes an implication of the above stated
observations. It states that all 1-streaks for $f$ of length at least
$3$ are saturated, i.e., they contribute significantly to $r_f(P)$.

\begin{lemma}\label{lem:basic_1-streak}
  Let $(P_n,f)$ be an $f$-reversible process. Then all 1-streaks of
  length $l\ge 3$ are saturated by all $c\in \Cf$.
\end{lemma}

The fourth threshold function of the last example demonstrates that
1-streaks of length $l< 3$ are not always saturated by all $c\in \Cf$.
These 1-streaks will be crucial in our analysis. Using induction,
Lemma~\ref{lem:splittable} and \ref{lem:basic_1-streak} yield that
$r_f(P) \ge \left\lceil n/2\right\rceil$ in case $f$ has a $1$-streak
of length $l\ge 2$. Thus, we put our focus on threshold functions,
where all $1$-streaks have length $0$ or $1$. We will show that each
such threshold function $f$ can be expressed as
$S_1\circ \ldots \circ S_k$ with
$\gamma(c_{S_i}) \ge \left\lceil \abs{S_i}/2\right\rceil$ for each
$S_i$ and each $c\in \Cf$. This will yield the lower bound. To
describe the partitioning of $f$ into subsequences, we introduce three
types of subsequences. Subsequences of type A are of the form
$(1,2^+,1)$, those of type B are of the form $(2^\ast)$ such that
nodes immediately before and after the sequence have threshold $1$ (if
they exist), and type $C$ of the form $(B,1^?,B)$, where the first B
is non-empty in case the $1$ occurs. Subsequences of type B and C can
be empty. If a subsequence of type C does not contain $1$, then it is
of type B. As an example consider the threshold function
$f=(2,2,1,2,2,1,2,2,2,1,2,1,2,1)$. Then
$f=(2,2) \circ (1,2,2,1) \circ (2,2,2) \circ (1,2,1) \circ (2,1)$,
i.e., $C=(2,1)$. For a configuration $c$ and a subsequence $S$ we
denote by $c_S$ the corresponding subsequence of $c$.

Before proving $r_f(P_n) \ge \left\lceil n/2\right\rceil$ we
provide properties of threshold functions having value $2$ for a
series of $l$ consecutive nodes. We will prove that in this case
$c_i=1$ for at least $\lfloor l/2 \rfloor$ of these nodes $v_i$, in some
cases even more.
  
  \begin{lemma}[Appendix~\ref{sec:proof-lemma-reflem_P1}]\label{lem:basic_P1}
    Let $(P_n,f)$ be an $f$-reversible process and $c\in \Cf$. The
    following properties hold:
    \begin{enumerate}
    \item\label{lem:basic_4} If $f_i=\ldots=f_{j}=2$ with $j\ge i$
      then $\gamma_{i,j}(c) \ge \lfloor l/2\rfloor$ with $l=j-i+1$. In
      particular, when $i=1$ or $j=n$ then
      $\gamma_{i,j}(c) \ge \lceil l/2\rceil$. If $f_i=2$ for all $i$
      then $\gamma(c)\ge\lceil (n+1)/2\rceil$.
    \item \label{lem:basic_6} If $f_1=1$, $f_2=\ldots=f_{i}=2$ with
      $i\ge 2$, and $f_{i+1}=1$ or $i=n$ then
      $\gamma_{1,i}(c) \ge \lceil i/2\rceil$. If $f_n=1$,
      $f_{i+1}=\ldots=f_{n-1}=2$ with $i\le n-2$, and $f_i=1$ or $i=0$
      then $\gamma_{i+1,n}(c) \ge \lceil (n-i)/2\rceil$.
    \item \label{lem:basic_7} If $f_{i+1}=f_{i+2}=2$,
      $f_{i}=f_{i+3}=1$, and $\gamma_{i,i+3}(c)=2$ then either
      $c_i=c_{i+1}=0$, $c_{i+2}=c_{i+3}=1$ or $c_i=c_{i+1}=1$,
      $c_{i+2}=c_{i+3}=0$, or $c_{i+1}=c_{i+2}=1$, $c_{i}=c_{i+3}=0$.
    \item \label{lem:basic_5} Let $f$ have no 1-streak of length $0$
      that is saturated by $c$. If $f_i=\ldots=f_{j}=2$ with
      $n>j\ge i>1$ and $f_{i-1}=f_{j+1}=1$ then
      $\gamma_{i-1,j+1}(c) \ge \lceil l/2\rceil + 1$ and
      $\gamma_{i,j}(c) \ge \lceil l/2\rceil$ with $l=j-i+1$. If $l$ is
      even and $l\ge 4$ then $\gamma_{i-1,j+1}(c) \ge l/2 +2$ and
      $c_{i-1}=c_{j+1}=1$.
    \end{enumerate}
  \end{lemma}

  This implies $r_1(P_n)=n$ and $r_2(P_n)=\lceil (n+1)/2\rceil$. With
  these results we can prove the existence of the above mentioned
  partitioning of $f$ into subsequences.
  
  \begin{lemma}\label{lem:basic_case}
    Let $(P_n,f)$ be an $f$-reversible process and $c\in \Cf$ such
    that $f$ has no 1-streak of length $0$ that is saturated by $c$.
    If $f$ has no 1-streak of length $2$ and if there are at least two
    nodes with threshold $1$ then $f$ can be uniquely partitioned into
    subsequences of type A,B, and C such that
    $f= (B \circ A)^+ \circ C$ and
    $\gamma(c_S) \ge \left\lceil \abs{S}/2\right\rceil$ for each
    occurring subsequence $S$.
  \end{lemma}
  \begin{proof}
    Clearly, $f$ can be uniquely expressed as
    $(B \circ A)^+ \circ \, C$ (subsequence of type C is the suffix
    following the last subsequence of type A). We claim that
    $\gamma(c_S) \ge \left\lceil \abs{S}/2\right\rceil$ for each
    $S\in \{A,B,C\}$. By Lemma~\ref{lem:basic_P1}.\ref{lem:basic_4} we
    have $\gamma(c_B) \ge \left\lceil \abs{B}/2\right\rceil$ for the
    leading/trailing subsequence of type B. By
    Lemma~\ref{lem:basic_P1}.\ref{lem:basic_5} we have
    $\gamma(c_A) \ge \left\lceil \abs{A}/2\right\rceil$ and
    $\gamma(c_B) \ge \left\lceil \abs{B}/2\right\rceil$ for all other
    subsequences of type $A$ or $B$.

    It remains to analyze $\gamma(c_C)$ in case $C$ contains a $1$. In
    case $C=B\circ 1$, i.e., $f_n=1$, the claim follows directly from
    Lemma~\ref{lem:basic_P1}.\ref{lem:basic_6}. Next, consider the
    case that $C=B_1\circ 1 \circ B_2$, where
    $B_i \not=\emptyset$ for $i=1,2$. If $\abs{B_2}$ is odd then
    $\gamma(c_C) \ge \left\lceil \abs{B_1}/2\right\rceil + \abs{1\circ
      B_2}/2 \ge \left\lceil \abs{C}/2\right\rceil$ by
    Lemma~\ref{lem:basic_P1}.\ref{lem:basic_4}. Assume that
    $\abs{B_2}$ is even. If $\abs{B_1}$ is odd then
    $\gamma(1\circ B_1 \circ 1)\ge (\abs{B_1}+1)/2 +1$ by
    Lemma~\ref{lem:basic_P1}.\ref{lem:basic_5}. Hence,
    $\gamma(B_1 \circ 1)\ge (\abs{B_1}+1)/2$ and therefore,
    $\gamma(c_C) \ge (\abs{B_1}+1+\abs{B_2})/2 = \abs{C}/2$. By the
    same lemma, if $\abs{B_1}$ is even and has length at least $4$
    then the label of the node of $C$ with threshold $1$ is $1$, thus
    $\gamma(c_C) \ge \abs{C}/2$.

    The only remaining case is that $f$ has a suffix $(1,2,2,1,B_2)$,
    with $\abs{B_2}\equiv 0(2)$. If $\abs{B_2}\ge 4$ then again the
    label of the node of $C$ with threshold $1$ is $1$ and we are
    done. Thus, consider the case that $f$ has a suffix
    $(1,2,2,1,2,2)$. Let $i$ be the index of the first node of $C$.
    Since $f$ has no 1-streak of length $2$ saturated by $c$ and
    $c\in \Cf$ we have either $c_i=0, c_{i+1}=1$ or
    $c_i=1, c_{i+1}=0$. The last three nodes cannot have labels
    $0,0,1$, because then $c$ would not converge (two neighboring
    nodes with threshold $2$ would have label $0$). This yields again
    that the label of the node of $C$ with threshold $1$ is $1$. Since
    the last node must have label $1$ this yields
    $\gamma(c_C) \ge \abs{C}/2$.
  \end{proof}

  Note that a similar result holds if $f$ is partitioned such that
  $f= C \circ (A \circ B)^+$, we just need to apply the last Lemma to
  the inverse of $f$. With this preparatory work we can prove our
  first main result.
  \begin{theorem}\label{theo:lowerBound}
    Let $(P_n,f)$ be an $f$-reversible process. Then
    $r_f(P_n) \ge \left\lceil n/2\right\rceil$.
  \end{theorem}
  \begin{proof}
    Let $f$ be a threshold function and $c\in \CSf$. We prove by
    induction that $\gamma(c)\ge \left\lceil n/2\right\rceil$.
    The case $n=2$ is trivial. Let $n\ge 3$. First assume that
    $f$ has a 1-streak of length $l$ and index $i$. If $l\ge 3$ then
    Lemma~\ref{lem:splittable} and induction yield
    \[r_f(P_n)= \gamma(c) = \gamma(c|_{\cev{P}_{i-1}}) +l +
      \gamma(c|_{\vec{P}_{l+i}})\ge \left\lceil
        \frac{i-1}{2}\right\rceil + l +\left\lceil
        \frac{n-l-i+1}{2}\right\rceil \ge \left\lceil
        \frac{n}{2}\right\rceil.\]

    Similarly, the case $l=2$ follows from
    Lemma~\ref{lem:splittable_2}. It remains to consider the case that
    each 1-streak has length at most $1$. If $f$ has a 1-streak of
    length $0$ that is saturated by $c$ then the theorem follows from
    Lemma~\ref{lem:splittable} and induction.

    If no node has threshold $1$ then the theorem follows from
    Lemma~\ref{lem:basic_P1}.\ref{lem:basic_4}. Consider the case that
    a single node $v_i$ has threshold $1$. If the first or last node
    has threshold $1$ the theorem follows from
    Lemma~\ref{lem:basic_P1}.\ref{lem:basic_6}. Since $c\in \Cf$ we
    cannot have $c_{i-1}=c_i=c_{i+1}=0$. Thus,
    $\gamma_{i-1,j+1}(c)\ge 1$. If the number of leading or trailing
    nodes with threshold $2$ is odd the theorem follows from
    Lemma~\ref{lem:basic_P1}.\ref{lem:basic_4}. If both numbers are
    even then $\gamma_{i-1,j+1}(c)\ge 1$ proves that the node with
    threshold $1$ has label $1$, which again yields the result. Hence,
    we can assume that there are at least two nodes with threshold
    $1$. Thus, the assumptions of Lemma~\ref{lem:basic_case} are
    satisfied and we have $f=S_1\circ \ldots \circ S_k$ with
    $\gamma_{S_i}(c) \ge \left\lceil\abs{S_i}/2\right\rceil$ for each
    $i$. Since the sum of the lengths of all $S_i$ equals $n$ the
    proof is complete.
  \end{proof}

  The bound of the theorem is sharp. Let $n\ge 2$ and
  $f=(2,\ldots,2,1)$ and $c(i)=1$ if $i$ is odd and $c(i)=0$
  otherwise, thus $\gamma(c)=\left\lceil n/2\right\rceil$. $c$
  converges to $\mathbb{1}$ in $n-1$ rounds. By
  Theorem~\ref{theo:lowerBound} the existence of a configuration
  $c\in C_f$ with $\gamma(c)=\left\lceil n/2\right\rceil$ implies
  $r_f(P_n)=\left\lceil n/2\right\rceil$.

  \begin{corollary}\label{cor:cor0}
      Let $(P_n,f)$ be an $f$-reversible process. If $f$ has a
      1-streak of length $l\ge 3$ then
      $r_f(P_n) \ge \left\lceil (n+l)/2\right\rceil$.
    \end{corollary}
    \begin{proof}
    Lemma~\ref{lem:basic_1-streak},
    Lemma~\ref{lem:splittable}, and Theorem~\ref{theo:lowerBound}
    yield
    \[r_f(P_n) =r_{f|_{\cev{P}_{i-1}}}(\cev{P}_{i-1}) + l +
    r_{f|_{\vec{P}_{l+i}}}(\vec{P}_{l+i}) \ge \left\lceil \frac{i-1}{2}\right\rceil + l
      +\left\lceil \frac{n-l-i+1}{2}\right\rceil \ge \frac{n+l}{2}.\]
  \end{proof}

  Corollary~\ref{cor:cor0} does not hold for $l=1,2$. Consider the case
  $n=3$ and $f=(2,1)$ (resp.\ $n=4$ and $f=(2,1,1,2)$). Then clearly
  $r_f(P_n)=1$ (resp.\ $r_f(P_n)=2$).
  
\section{Threshold Functions with $r_f(P_n)=n$}
\label{sec:case-r_fp=n}

In this section we characterize threshold functions $f$, which only
have the trivial conversion set $V$, i.e., $r_f(P_n)=n$. Using this
characterization, we will construct a DFA with seven states that
accepts these threshold functions. The number of accepting input
strings of length $n$ of this DFA can be computed using dynamic
programming in linear time.
\begin{definition}
Let ${\cal F}_n$ be the set of threshold
functions $f$ of length $n$ with $f_i\in \{1,2\}$ satisfying the
following two properties:
  \begin{enumerate}
  \item There is no $i$ such that $f_i=f_{i+1}=f_{i+2}=2$, or
    $f_i=f_{i+2}=2, f_{i+1}=1$, or $f_i=f_{i+3}=2$, $f_{i+1}=f_{i+2}=1$.
  \item If $f_1=1$ then $f_2=1$ and if $f_n=1$ then $f_{n-1}=1$.%
  \end{enumerate}  
\end{definition}
  \begin{theorem}\label{theo:bound}
    Let $(P_n,f)$ be an $f$-reversible process. Then $r_f(P_n)=n$ if
    and only if $f\in {\cal F}_n$.
  \end{theorem}
  \begin{proof}
    Let $f$ be a threshold function of $P$ with $r_f(P)=n$. Assume
    there exists $i\in \{1,\ldots,n-2\}$ such that
    $f_i=f_{i+1}=f_{i+2}=2$. Then configuration $C^{i+1}$ leads in one
    round to configuration $\mathbb{1}$. Thus, $r_f(P)<n$, a
    contradiction. Assume there exist $i,j$ with $2 \le j-i \le 3$,
    $f_i=f_j=2$ and $f_{i+1}= \ldots =f_{j-1}=1$. Let
    $S_{i,j}=\{i+1,\ldots,j-1\}$. The inverse of configuration
    $c_{S_{i,j}}$ reaches in one round configuration $\mathbb{1}$, a
    contradiction. Suppose $f_1=1$ and $f_2=2$. Then the initial
    configuration $C^{1}$ leads in one round to configuration
    $\mathbb{1}$, a contradiction. Similarly, if $f_n=1$ then
    $f_{n-1}=1$. Thus, $f\in {\cal F}_n$.

    Next, consider $f\in {\cal F}_n$. We will prove by induction on $n$
    that for each $c\in C_f$ we have $\gamma(c)=n$, i.e.,
    $c=\mathbb{1}$. For $n\le5$ the statement can be easily verified
    manually. Let $n\ge 6$ and $c\in C_f$. If $f=\mathbb{1}$ then
    clearly $c=\mathbb{1}$. Consider $f\not=\mathbb{1}$. We claim that
    if $f_i=2$ then $c_j=1$ for all $j\ge i$. Assume that this claim
    is false. Choose $i$ maximal such that the claim is false. Clearly
    $i<n$, since if $i=n$ then $c_i=1$ by
    Lemma~\ref{lem:basic_P1}.\ref{lem:basic_4}. If $i=n-1$ then
    $f_{i+1}=f_n=2$ since $f\in {\cal F}_n$. Again
    Lemma~\ref{lem:basic_P1}.\ref{lem:basic_4} yields $c_{n}=1$. Thus,
    $c_{n-1}=0$ by the choice of $i$. Then $c_{n-2}=c_{n-3}=c_{n-4}=1$
    since $f\in {\cal F}_n$. But this implies $c_{n-1}=1$ since
    $c\in \Cf$. Hence, $i+1<n$. Assume that $f_{i+1}=2$. Then
    $f_{i+2}=1$ since $f\in {\cal F}_n$ and $c_{j}=1$ for $j\ge i+1$
    by the choice of $i$. Clearly,
    $f|_{\cev{P}_{i+1}}\in {\cal F}_{i+1}$ and
    $c|_{\cev{P}_{i+1}}\in C_{f|_{\cev{P}_{i+1}}}\!.$ By induction
    $\gamma(c|_{\cev{P}_{i+1}}) =i+1$ and hence $c_i=1$. It remains to
    consider the case $f_{i+1}=1$. Again $f_{i+2}=1$ since
    $f\in {\cal F}_n$. Since $c\in C_f$ we have $c_{i+1}=c_{i+2}$. It
    is easy to see that this yields $c_{i+1}=c_{i+2}=1$ and therefore,
    $c_i=1$. Furthermore, $c_j=1$ for $j\ge i+1$ by the choice of $i$.
    This completes the proof of the claim. Similarly we prove that if
    $f_i=2$ then $c_j=1$ for all $j\le i$. Hence, since
    $f\not=\mathbb{1}$ we have proved that $c=\mathbb{1}$.
  \end{proof}

\begin{figure}[h]
  \hfill
  \includegraphics{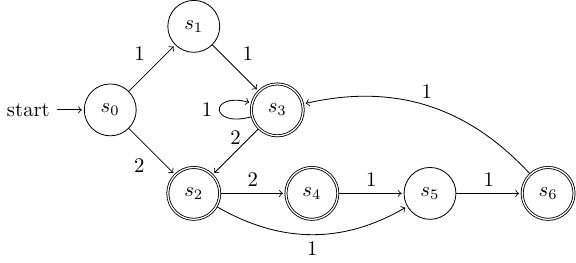}
  \hfill\null 
   \caption{A DFA to recognize threshold functions with $r_f(P ) = n$
   for $n \ge 2$.}\label{fig:DFA}
\end{figure}

  \begin{theorem}\label{theo:time_1}
    The number of threshold functions $f$ such that $(P_n,f)$ is an
    $f$-reversible process with $r_f(P_n)=n$ can be computed in $O(n)$
    time.
  \end{theorem}
  \begin{proof}
    By Theorem~\ref{theo:bound} it suffices to compute
    $\abs{{\cal F}_n}$. Fig.~\ref{fig:DFA} depicts a DFA to recognize
    ${\cal F}_n$, it is independent of $n$. The number of accepting
    input strings of length $n$ can be easily computed using dynamic
    programming in linear time $O(n)$.
  \end{proof}

This DFA also allows to list all threshold functions $f$ with
$r_f(P)=n$ in time $O(n+I)$ where $I$ is the number of such threshold
functions. It is possible to decide in linear time whether $r_f(P)=n$
for a given threshold functions $f$.

\section{Threshold Functions with $r_f(P_n)=\left\lceil n/2\right\rceil$}
\label{sec:case-r_fp=n_2}

The characterization of threshold functions $f$ supporting a labeling
$c\in \Cfp{\left\lceil n/2\right\rceil}$ depends on the parity of $n$.
In Definition~\ref{def:East} and~\ref{def:Oast} we define for even
(resp.\ odd) $n$ a set ${\cal E}^e_n$ (resp.\ ${\cal E}^o_n$) such
that $r_f(P_n)=\left\lceil n/2\right\rceil$ if and only if
$f\in {\cal E}^e_n$ (resp.\ ${\cal E}^o_n$).

\subsection{The Case $n\equiv 0(2)$} \label{sec:case-nequiv-02} If $n$
is even, then $f$ has by Lemma~\ref{lem:splittable} no 1-streak of
length $l\ge 1$ saturated by any $c\in \Cfp{n/2}$. We first
characterize threshold functions without a saturated 1-streak by
$c\in \Cfp{n/2}$ of length~$0$. Let ${\cal E}_2=\{F^{1,2}, F^{2,2}\}$
and ${\cal E}_n= \{F^{1,n}, F^{n,n},F^{3,n},F^{n-2,n}\}$ for $n\ge 4$.
If $n\equiv 0(4)$ then let
${\cal E}^\ast_n= {\cal E}_n \cup \{F^{n/2,n/2}\circ F^{1,n/2}\}$ else
let ${\cal E}^\ast_n= {\cal E}_n$. For each
$F^{i,n}\in {\cal E}^\ast_n$ we define a configuration $c^{F^{i,n}}$
as follows: $(c^{F^{i,n}})_i=0$ and for each end node of $P_n$ with
threshold $2$ we start assigning the labels $1$ and $0$ alternating
until we reach $v_i$. It is easy to verify that
$c^{F^{i,n}}\in \Cfp{n/2}$. Thus, $r_{F^{i,n}}(P_n)=n/2$. Furthermore,
we have
$c^{F^{n/2,n/n}}\circ c^{F^{1,n/n}}\in {\cal C}_{F^{n/2,n/2}\circ
  F^{1,n/2}}(n/2)$. Hence, $r_{f}(P_n)=n/2$ for all
$f\in {\cal E}^\ast_n$.

  \begin{lemma}[Appendix~\ref{sec:proof-lemma-even_case}]\label{lem:even_case}
    Let $(P_n,f)$ be an $f$-reversible process with $n$ even. Then $f$
    has no 1-streak of length $0$ saturated by a $c\in \Cfp{n/2}$ and
    no 1-streak of length $2$ if and only if $f\in {\cal E}_n$.
  \end{lemma}

  Lemma~\ref{lem:even_case} shows that if $n\ge 4$ is even, there are
  only four different threshold functions $f$ with $r_f(P_n)=n/2$
  provided there are no saturated 1-streaks of length $0$ and
  1-streaks of length $2$. Next, we generalize this characterization by
  omitting the second requirement.

\begin{theorem}[Appendix~\ref{sec:proof-theo:even_case_2}]\label{theo:even_case_2}
  Let $(P_n,f)$ be an $f$-reversible process with $n$ even. Then
  $f\in {\cal E}_n^\ast$ if and only if no 1-streak of $f$ of length
  $0$ is saturated by a $c\in \Cfp{n/2}$.
\end{theorem}  

\begin{definition}\label{def:East}
  Let $(P_n,f)$ be an $f$-reversible process with $n$ even. Then
  $f\in {\cal F}^e_n$ if and only if $P$ can be partitioned into
  sub\-paths $P^1,\ldots,P^k$ such that each $P^i$ has even length
  $l_i$, $f_{P^i}\in {\cal E}_{l_i}^\ast$, and for $i<k$ the following
  two conditions hold:
  \begin{enumerate}
  \item The last value of $f_{P^i}$ coincides with the first value of
    $f_{P^{i+1}}$, and
  \item if the last value of $f_{P^i}$ equals $1$ then $l_i=l_{i+1}$.
  \end{enumerate}
\end{definition}

Theorems~\ref{theo:lowerBound} and~\ref{theo:even_case_2} together
with Lemma~\ref{lem:splittable} yield the following result.

\begin{theorem}\label{cor:cor1}
  Let $(P_n,f)$ be an $f$-reversible process with $n$ even. Then
  $r_f(P_n)=n/2$ if and only if $f\in {\cal F}^e_n$.
  \end{theorem}
\begin{proof}
  Let $f\in {\cal F}^e_n$. By definition we have
  $f_{P^i}\in {\cal E}_{f_{P^i}}(l_i/2)$. For $i=1,\ldots,k$ let $c^i$
  be the configuration constructed at the beginning of
  Section~\ref{sec:case-nequiv-02} with
  $c^i\in {\cal C}_{f_{P^i}}(l_i/2)$. Note that each $c^i$ converges
  independently to $\mathbb{1}$. Then
  $c^1\circ \ldots \circ c^k \in \Cf(n/2)$. Hence, $r_f(P_n)=n/2$.

  Next, let $f$ be a threshold function of $P_n$ with $r_f(P_n)=n/2$
  and let $c\in \Cfp{n/2}$. If no 1-streak of $f$ length $0$ is
  saturated by $c$ then Theorem~\ref{theo:even_case_2} yields
  $f\in {\cal E}^\ast_n$, i.e., $f\in {\cal F}^e_n$. Otherwise let $i$
  be an index of a 1-streak of length $0$ saturated by $c$. Then
  \[n/2=r_f(P_n) =r_{f|_{\cev{P}_{i}}}(\cev{P}_{i}) +
    r_{f|_{\vec{P}_{i+1}}}(\vec{P}_{i+1})\ge \left\lceil
      i/2\right\rceil + \left\lceil (n-i)/2\right\rceil\] by
  Lemma~\ref{lem:splittable} and Theorem~\ref{theo:lowerBound}. This
  yields that $i$ is even. Thus, the indices of 1-streaks of $f$ of
  length $0$ that are saturated by $c$ divide $P_n$ in to even-length
  sub\-paths $P^1,\ldots,P^k$. Clearly,
  $c_{P^i} \in \Cfp{\abs{P^i}/2}$. Note that no $P^i$ has a 1-streak
  of $f_{P^i}$ of length $0$ that is saturated by $c_{P^i}$. Hence,
  Theorem~\ref{theo:even_case_2} can be applied to $P^i$. Let
  $l_i=\abs{P^i}$. This yields $f|_{P^i}\in {\cal E}^\ast_{l_i}$.

  Let $i<k$. Assume the last value of $f|_{P^i}$ does not coincide with
  the first value of $f|_{P^{i+1}}$. This is only possible if either
  $f|_{P^i}=F^{l_i,l_i}$ and $f_{P^{i+1}}=F^{3,l_{i+1}}$ or
  $f_{P^{i+1}}=F^{l_{i+1}-2,l_{i+1}}$ and secondly,
  $f_{P^i}=F^{3,l_{i}}$ or $f_{P_{i}}=F^{l_{i}-2,l_{i}}$ and
  $f_{P^{i+1}}=F^{1,l_{i+1}}$. In both cases $c$ does not converge to
  $\mathbb{1}$. This proves $f\in {\cal F}^e_n$.
\end{proof}

The last theorem leads to an algorithm constructing all threshold
functions with $r_f(P_n)=n/2$ for even $n$. First, determine all
compositions of $n$, where each term is even. For each composition
replace each term $t$ by a threshold function of length $t$ according
to Definition~\ref{def:East} such that the stated constraints are
satisfied.

Theorem~\ref{theo:lowerBound} yields that
$r^c_f(P_n)\ge \left\lceil n/2\right\rceil$. Surprisingly, there are
threshold functions with $r^c_f(P_n)= \left\lceil n/2\right\rceil$. It
is not difficult to characterize those. E.g., if $n\equiv 0 (4)$ then
these are the functions $f=(1,2,2,1)^+$ and $f=(2,1,1,2)^+$ and if
$n\equiv 2 (4)$ the functions $f=(1,2,2,1)^+\circ(1,2)$ and
$f=(2,1,1,2)^+ \circ (2,1)$.

\subsection{The Case $n\equiv 1 (2)$}
\label{sec:case-nequiv-1}
If $n$ is odd, Lemma~\ref{lem:splittable} does not exclude that $f$ has
a 1-streak of length $1$ saturated by a configuration
$c\in \Cf(\left\lceil n/2\right\rceil)$. In this case the index $i$
must be odd. A 1-streak of $f$ of length $2$ cannot be saturated by
any $c\in \Cf(\left\lceil n/2\right\rceil)$, but there still can be a
unsaturated 1-streaks of length $2$. For the characterization we
consider two cases, either $f$ has 1-streak of length $2$ or it does
not. In both cases we show that $f$ can be decomposed into smaller
segments, such that if $c\in \Cf(\left\lceil n/2\right\rceil)$, then
$c$ restricted to each segment converges independently to
$\mathbb{1}$.

A simple example demonstrates this approach. Let $n_1+n_2=n-1$ with
$n_k> 0$ and $n_k\equiv 0(2)$ for $k=1,2$, $P_{n_k}$ a path with $n_k$
nodes, and $f^k$ a threshold function for $P_{n_k}$ with
$f^1_{n_1}=f^2_1=2$. If $c^k$ is a labeling for $P_{n_k}$ converging
for $f^k$ to $\mathbb{1}$ with $\gamma(c^k)=n_k/2$ then
$c=c^1 \circ 1 \circ c^2$ is a configuration for $P_n$ that converges
for $f=f^1 \circ 1 \circ f^2$ and for $f=f^1 \circ 2 \circ f^2$ to
$\mathbb{1}$ and $\gamma(c)=\left\lceil n/2\right\rceil$. In the first
(resp.\ second) case $f$ has a 1-streak of length $1$ (resp.\ $0$)
that is saturated by $c$. A similar statement can be made if $n_1=0$
or $n_2=0$. It is worth mentioning that there are threshold functions
$f$ with $r_f(P)=\left\lceil n/2\right\rceil$ that do not follow this
pattern. For $n=7$ and $f=(2,1,1,2,1,2,2)$ the configuration
$(1,0,0,1,1,0,1)$ converges, i.e., $r_f(P)=4$, but $f$ has no 1-streak
saturated by $c$.

To characterize the above mentioned segments, we introduce the set
${\cal O}_n^\ast$. For $n\equiv 1 (2)$ let ${\cal O}_n^\ast$ be the
set of the following threshold functions:
\begin{itemize}
\item $F^n$ and $F^{k,n}$, $k=1,\ldots,n$
\item $F^{k,k} \circ F^{n-k,n-k}$, $k=5,7,\ldots,n-2$ and
  $F^{1,k}\circ F^{1,n-k}$, $k=2,4,\ldots,n-1$
\item $F^{k,k} \circ F^{n-k-2,n-k}$, $k=5,7,\ldots,n-4$ and
  $F^{3,k} \circ F^{1,n-k}$, $k=4,6,\ldots,n-1$
\item $F^{1,3}\circ F^{1,n-3}$ and $F^{n-3,n-3} \circ F^{3,3}$ 
\item
  $F^{k,k} \circ 1 \circ F^{k,n-k-1}, F^{n-2k,n-k-1}\circ 1 \circ
  F^{1,k}$, $k=2,4,\ldots,2\left\lfloor (n-1)/4\right\rfloor$.
\end{itemize}

It is easy to construct for each $f\in {\cal O}_n^\ast$ a
configuration $c$ such that $c\in \Cf(\left\lceil n/2\right\rceil)$.
E.g., for $f=F^{1,3}\circ F^{1,n-3}$ use $c=(1,1,0,0,1,0,1,\ldots)$.
 
\begin{lemma}[Appendix~\ref{sec:proof-theo:odd_case}] \label{theo:odd_case}
  Let $(P_n,f)$ be an $f$-reversible process with $n$ odd. If $f$ has
  no 1-streak of length $0$ or $1$ saturated by a
  $c\in \Cf(\left\lceil n/2\right\rceil)$ then $f\in {\cal O}_n^\ast$.
\end{lemma}

\begin{definition}\label {def:Oast}
  Let $(P_n,f)$ be an $f$-reversible process with $n$ odd. Then
  $f \in {\cal F}^o_n$ if and only if there exists
  $1\le i \le j \le n$ with $i,j$ odd such that
  $f_{\cev{P}_{i-1}}\in {\cal F}^e_{i-1}$, $f_{\vec{P}_{j+1}} \in {\cal
    F}^e_{n-j}$ and either $i=j$ and $f_{{P}_{i,j}}\in \{1,2\}$ or
  $j>i+1$ and $f_{{P}_{i,j}}\in {\cal O}^\ast_{j-i+1}$ such that
  $f=f_{\cev{P}_{i-1}} \circ f_{{P}_{i,j}} \circ f_{\vec{P}_{j+1}}$
  and the following two conditions hold:
  \begin{enumerate}
  \item If $i>1$ then $(f_{\cev{P}_{i-1}})_{i-1}=2$, if $j< n$ then
    $(f_{\vec{P}_{j+1}})_1=2$, and
  \item if $1<i<j$ then $(f_{{P}_{i,j}})_1=2$, and if $i<j<n$ then
    $(f_{{P}_{i,j}})_{j-i+1}=2$.
  \end{enumerate}
\end{definition}

The last two conditions request that the values of the threshold
functions at the junctions are equal to $2$. This guarantees that each
of $f_{\cev{P}_{i-1}}, f_{\vec{P}_{j+1}},$ and $f_{\cev{P}_{i-1}}$
converge independently. The following theorem gives a complete
characterization of all threshold functions $f$ with
$r_f(P_n)=\left\lceil n/2\right\rceil$ in case $n$ is odd.

\begin{theorem}\label{theo:n_half_fin}
  Let $(P_n,f)$ be an $f$-reversible process with $n$ odd. Then
  $r_f(P_n)=\left\lceil n/2\right\rceil$ if and only if
  $f\in {\cal F}^o_n$.
\end{theorem}%

\begin{proof}
  Let $f\in {\cal F}^o_n$. By definition there exists
  $c_{\cev{P}_{i-1}}\in {\cal C}_{f_{\cev{P}_{i-1}}}((i-1)/2)$ and
  $c_{\vec{P}_{j+1}}\in {\cal C}_{f_{\vec{P}_{j+1}}}((n-i)/2)$. If
  $f_{{P}_{i,j}}\in \{1,2\}$ then
  $c_{\cev{P}_{i-1}}\circ 1 \circ c_{\vec{P}_{j+1}}\in \Cf(\left\lceil
    n/2\right\rceil)$, i.e., $r_f(P)=\left\lceil n/2\right\rceil$. If
  $f_{{P}_{i,j}}\in {\cal O}^\ast_{j-i+1}$ then there exists
  $c_{i,j}\in {\cal C}_{{P}_{i,j}}(\left\lceil(j-i+1/2\right\rceil)$.
  Thus,
  $c_{\cev{P}_{i-1}}\circ c_{i,j} \circ c_{\vec{P}_{j+1}}\in
  \Cf(\left\lceil n/2\right\rceil)$, i.e., $r_f(P)=n/2$.

  Next, consider $f$ with $r_f(P_n)=\left\lceil n/2\right\rceil$
  and $c\in \Cfp{\left\lceil n/2\right\rceil}$. We have two cases.

  \noindent
  \textbf{Case 1}: There exists a 1-streak of length $0$ saturated by
  $c$: Let $i_1<i_2<\dots<i_k$ be the indices of 1-streaks of length
  $0$ saturated by $c$. Assume that $i_j+1=i_{j+1}$ for some $j$. Then
  $f_{i_j}=f_{i_{j}+1}=f_{i_{j}+2}=2$ and
  $c_{i_j}=c_{i_{j}+1}=c_{i_{j}+2}=1$. Hence,
  $f=f|_{\cev{P}_{i_j}}\circ 2 \circ f|_{\vec{P}_{i_j+2}}$. Also
  $c|_{\cev{P}_{i_j}}$ (resp.\ $c|_{\vec{P}_{i_j+2}}$) converges on
  $\cev{P}_{i_j}$ (resp.\ $\vec{P}_{i_j+2}$) to $\mathbb{1}$. Then
  Theorem~\ref{theo:lowerBound} implies that $\abs{\cev{P}_{i_j}}$ and
  $\abs{\vec{P}_{i_j+2}}$ are even,
  $r_{f|_{\cev{P}_{i_j}}}(\cev{P}_{i_j})=\abs{\cev{P}_{i_j}}/2$, and
  $r_{f|_{\vec{P}_{i_j+2}}}(\vec{P}_{i_j+2})=\abs{\vec{P}_{i_j+2}}/2$.
  Hence, $f|_{\cev{P}_{i_j}}\in {\cal F}^e_{i_j}$ and
  $f|_{\vec{P}_{i_j+2}}\in {\cal F}^e_{n-i_j-1}$ by
  Theorem~\ref{cor:cor1}. Therefore, $f\in {\cal F}^o_n$.

  Next, assume that $i_j+1\not= i_{j+1}$ for $j=1,\ldots,k-1$. Thus,
  $i_j+ 2 < i_{j+1}$. The indices $i_1,\dots,i_k$ divide $f$ into
  $k+1$ segments of the form $(f_{i_j+1},\ldots,f_{i_{j+1}})$, inner
  segments are of length at least $3$. Each segment converges to
  $\mathbb{1}$ and no segment has a 1-streak of length $0$ saturated
  by $c$. By Theorem~\ref{theo:lowerBound} all but one have even
  length. Let $i_j+1$ be the first index of the odd length segment.
  Then
  $f=f_{\cev{P}_{i_j}}\circ f_{P_{i_j+1,i_{j+1}}} \circ
  f_{\vec{P}_{i_{j+1}+1}}$. All even length segments satisfy the
  conditions of Theorem~\ref{theo:even_case_2}. Thus,
  $f|_{\cev{P}_{i_j}}\in {\cal F}^e_{i_j}$,
  $f|_{\vec{P}_{i_{j+1}+1}}\in {\cal F}^e_{n-i_{j+1}}$ by
  Theorem~\ref{cor:cor1}, i.e., $f\in {\cal F}^o_n$.

  Next, we consider the segment $f_{P_{i_j+1,i_{j+1}}}$ of odd length
  $l^o=i_{j+1}-i_j$. If it has no 1-streak of length $1$ saturated by
  $c$ then we apply Lemma~\ref{theo:odd_case}, i.e.,
  $f_{P_{i_j+1,i_{j+1}}}\in {\cal O}_{l^o}^\ast$ and thus,
  $f\in {\cal F}^o_n$. So assume that the segment has a 1-streak of
  index $i$ and length $1$ saturated by $c$. Then
  Lemma~\ref{lem:splittable} and Theorem~\ref{theo:lowerBound} yield
  \[(l^o+1)/2 \ge \left\lceil (i-1)/2 \right\rceil + 1 +\left\lceil
    (l^o-i)/2\right\rceil.\] This is only possible if $i\equiv 1 (2)$.
  Suppose that there exists another 1-streak of length $1$ saturated
  by $c$. Let $j > i+1$ be smallest index of such a 1-streak. Then $c$
  converges on each of the segments $\cev{P}_{i-1}$, $P_{i+1,j-1}$,
  $\vec{P}_{j+1}$ to $\mathbb{1}$. $P_{i+1,j-1}$ consists of an odd
  number of nodes and $f_{P_{i+1,j-1}}$ has no 1-streak of length $0$
  or $1$ saturated by $c_{P_{i+1,j-1}}$. Thus,
  $\gamma(c_{P_{i+1,j-1}}) \ge (j-i+1)/2 +1$ by
  Lemma~\ref{lem:basic_P1}.\ref{lem:basic_5}. Now
  Theorem~\ref{theo:lowerBound} leads to a contradiction. Hence, there
  is at most one 1-streak of length $1$ that is saturated by $c$.
  Applying Lemma~\ref{lem:splittable} to $\cev{P}_{i-1}$ and
  $\vec{P}_{i+1}$ yields
  $f=f|_{\cev{P}_{i-1}}\circ 1 \circ f|_{\vec{P}_{i+1}}$, i.e.,
  Theorem~\ref{cor:cor1} yields $f\in {\cal F}^o_n$.

  \textbf{Case 2}: There exists no 1-streak of length $0$ saturated by
  $c$.

  If there exists a 1-streak of index $i$ and length $1$ saturated by
  $c$ then we repeat the argument for the single segment of odd length
  in the first case proving that there exists only one such 1-streak.
  Again $f=f|_{\cev{P}_{i-1}}\circ 1 \circ f|_{\vec{P}_{i+1}}$ and
  Theorem~\ref{cor:cor1} can be applied. If there exists no 1-streak
  length $1$ saturated by $c$ then we can apply
  Lemma~\ref{theo:odd_case} to prove $f\in {\cal O}_n^\ast$, i.e.,
  $f\in {\cal F}^o_n$. This completes the proof.
\end{proof}

\section{Conclusion}
\label{sec:conclusion}
In this paper we analyzed $f$-reversible processes $(P_n,f)$ with
$P_n$ a path of length $n$. We proved
$r_f(P_n)\ge \left\lceil n/2\right\rceil$ and gave a characterization
of threshold functions with $r_f(P_n)=\left\lceil n/2\right\rceil$ or
$r_f(P_n)=n$. It would be interesting to prove results similar to
those of this paper for trees or for graphs of small pathwidth.
However the lower bound of Theorem~\ref{theo:lowerBound} does not even
hold for simple trees. Consider a star graph $S_n$, where each of the
$n-1$ leaves has threshold $1$ and the inner node has threshold $n-1$.
Then $r_f(S_n)=2$ for all $n\ge 3$. It is an open question, whether
the results hold for trees in which all nodes have threshold $1$ or
$2$.

\bibliographystyle{splncs04}
\bibliography{conversion}

\appendix
\section{Appendix}

\subsection{Proofs of Section~\ref{sec:basic-properties-f}}

\subsubsection{Proof of Lemma~\ref{lem:splittable}}
  \label{sec:proof-lemma-splittable}
\begin{proof}
  To simplify notation we introduce two pseudo nodes $v_0$ and
  $v_{n+1}$ that are adjacent to the end nodes of $P_n$, both have
  threshold $2$. Furthermore, $\cev{P}_{{0}}$ and $\vec{P}_{{n+1}}$
  denote the empty graph. First we consider the case $l\ge 1$. Clearly
  nodes $v_{i-1},\ldots,v_{l+i}$ never change their labels. Hence,
  $c|_{\cev{P}_{i-1}}\in {\cal C}_{f|_{\cev{P}_{i-1}}}(\cev{P}_{i-1})$
  and
  $c|_{\vec{P}_{l+i}}\in {\cal
    C}_{f|_{\vec{P}_{l+i}}}(\vec{P}_{l+i})$. Therefore,
  $\gamma(c|_{\cev{P}_{i-1}})\ge
  r_{f|_{\cev{P}_{i-1}}}(\cev{P}_{i-1})$ and
  $\gamma(c|_{\vec{P}_{j+1}})\ge
  r_{f|_{\vec{P}_{j+1}}}(\vec{P}_{j+1})$. This yields
  $r_f(P) \ge r_{f|_{\cev{P}_{i-1}}}(\cev{P}_{i-1}) + l +
  r_{f|_{\vec{P}_{l+i}}}(\vec{P}_{l+i})$. Let $c^1$ (resp.\ $c^2$) be
  a configuration of $\cev{P}_{i-1}$ (resp.\ $\vec{P}_{l+i}$) that
  converges to $\mathbb{1}$ with $\gamma(c^1)=r_f(\cev{P}_{i-1})$
  (resp.\ $\gamma(c^2)=r_f(\vec{P}_{l+i})$). Define
  $c'=c^1\circ 1\circ \ldots \circ 1\circ c^2$. Then $c'$ is a
  configuration of $P$. If $i>1$ then $c^1_{i-1}=1$. Similarly, if
  $j<n$ then $c^2_{j+1}=1$. Thus, clearly we have $c'\in C_f$ and
  hence
  $r_{f|_{\cev{P}_{i-1}}}(\cev{P}_{i-1}) + l +
  r_{f|_{\vec{P}_{j+1}}}(\vec{P}_{j+1}) \ge r_f(P)$.

  If $l=0$ and $c_i=c_{i+1}=1$ then nodes $v_i$ and $v_{i+1}$ never
  change their labels. Thus, ${c|_{\cev{P}_{i}}}$ and
  ${c|_{\vec{P}_{i+1}}}$ converge to $\mathbb{1}$. Hence,
  $r_f(P) \ge r_{f|_{\cev{P}_{i}}}(\cev{P}_{i}) +
  r_{f|_{\vec{P}_{i+1}}}(\vec{P}_{i+1})$. Let $c^1$ (resp.\ $c^2$) be
  a configuration of $\cev{P}_{i}$ (resp.\ $\vec{P}_{i+1}$) that
  converges to $\mathbb{1}$ with $\gamma(c^1)=r_f(\cev{P}_{i})$
  (resp.\ $\gamma(c^2)=r_f(\vec{P}_{i+1})$). Then $c^1(i)=c^2(1)=1$.
  Let $c'=c^1\circ c^2$. Then $c'$ is a configuration of $P$. If $i>1$
  then $c^1_{i-1}=1$. Similarly, if $j<n$ then $c^2_{j+1}=1$. Thus,
  clearly we have $c'\in C_f$. Hence
  $r_{f|_{\cev{P}_{i}}}(\cev{P}_{i}) +
  r_{f|_{\vec{P}_{i+1}}}(\vec{P}_{i+1}) \ge r_f(P)$.
\end{proof}

\subsubsection{Proof of Lemma~\ref{lem:splittable_2}}
  \label{sec:proof-lemma-splittable_2}
\begin{proof}
  Let $c\in \CSf$. Since $f_i=f_{i+1}=1$ we have $c_i=c_{i+1}$ in all
  future rounds. Assume $i=1$. Then node $v_{3}$ exists and $f_{3}=2$.
  Clearly $v_1,v_2$ and $v_3$ all have the same label and will never
  change this label. Since $c$ converges to $\mathbb{1}$ we have
  $c_1=c_2=c_3=1$. This proves the result. The case $i+1=n$ is proved
  similarly. Thus, both nodes $v_{i-1},v_{i+2}$ exist. Since
  $c_{i}=c_{i+1}$ in all future rounds, also $c_{i-1}=c_{i+2}$ in all
  future rounds. Thus, ${c|_{\cev{P}_{i}}}$ and ${c|_{\vec{P}_{i+1}}}$
  converge to $\mathbb{1}$ and
  $r_f(P_n) = \gamma(c_{\cev{P}_{i}}) + \gamma(c_{\vec{P}_{i+1}}) \ge
  r_{f|_{\cev{P}_{i}}}(\cev{P}_{i}) +
  r_{f|_{\vec{P}_{i+1}}}(\vec{P}_{i+1})$.
\end{proof}

\subsubsection{Proof of Lemma~\ref{lem:basic_P1}}
\label{sec:proof-lemma-reflem_P1}

\begin{proof}
  \begin{enumerate}
  \item If $f_k=f_{k+1}=2$ for some $k$ then $c_k=1$ or $c_{k+1}=1$.
    Thus, $\gamma_{i,j}(c) \ge \lfloor l/2\rfloor$. If $f_1=2$ (resp.\
    $f_n=2$) then $c_1=1$ (resp.\ $c_n=1$) since otherwise node $v_1$
    (resp.\ $v_n$ could never change its label to $1$. Thus, in this
    case $\gamma_{i,j}(c) \ge \lceil l/2\rceil$. If $f_i=2$ for all
    $i$ then $c_1=c_n=1$, thus $\gamma(c)=\lceil (n+1)/2\rceil$.
  \item If $i=n$ then $c_n=1$. If $c_1=0$ then $c_2=1$. Thus,
    $\gamma_{2,i}(c) \ge \lceil (i+1)/2\rceil$ as above. Hence,
    $\gamma_{1,i}(c) \ge \lceil i/2\rceil$. Let $f_{i+1}=1$. Then
    $\gamma_{2,i}(c) \ge \lceil (i-1)/2\rceil$ by the first property.
    If $c_1=1$ or $c_n=1$ the statement clearly holds. If $c_1=c_n=0$
    then $c_2=c_{n-1}=1$ and the result follows as before.
  \item Since $\gamma_{i,i+3}(c)=2$ there are $6$ possible
    configurations. By the assumption it is impossible that $c_{i+1}$
    and $c_{i+2}$ are both $0$. For the same reason the configurations
    $c_i=c_{i+2}=0$, $c_{i+1}=c_{i+3}=1$ and $c_i=c_{i+2}=1$,
    $c_{i+1}=c_{i+3}=0$ are impossible.
  \item Let $l=1$. Assume that $c_i=0$. For $v_i$ to change its label to
  $1$ we must have in some round $c_{i-1}=c_{i+1}=1$. But then in the
  following rounds the labels of $v_{i-1},v_i, v_{i+1}$ toggle between
  $1,0,1$ and $0,1,0$. Since $c \in \Cf$, this is impossible, thus,
  $c_i=1$. For the same reason it is impossible that in this case
  $c_{i-1}=c_{i+1}=0$. Hence, $\gamma_{i-1,i+1}(c) \ge 2$. Next,
  consider the case $l=2$. By
  Lemma~\ref{lem:basic_P1}.\ref{lem:basic_4} $\gamma_{i,j}(c)\ge 1$.
  Thus we can assume that $c_i=1$ and $c_{i+1}=0$. If
  $c_{i-1}=c_{i+2}=0$ then in one round we would have $c_i=c_{i+1}=0$,
  a contradiction to Lemma~\ref{lem:basic_P1}.\ref{lem:basic_4}.

  Finally, we consider the case $l\ge 3$. By
  Lemma~\ref{lem:basic_P1}.\ref{lem:basic_4} we have
  $\gamma_{i,j}(c)\ge \lfloor l/2\rfloor$. First assume $l$ is even,
  i.e., $\lfloor l/2\rfloor=l/2$. If $\gamma_{i,j}(c)> l/2$ then
  $\gamma_{i-1,j+1}(c) \ge \gamma_{i,j}(c)\ge 1+ l/2 = \lceil
  (l+2)/2\rceil$ and we are done. So assume $\gamma_{i,j}(c)= l/2$.
  WLOG $c_j=0$. Then $c_{j+1}=1$, otherwise in one round
  $c_{j-1}=c_j=0$ which contradicts property~\ref{lem:basic_4}. Thus,
  $\gamma_{i-1,j+1}(c) \ge 1 + l/2 = \lceil l/2\rceil +1$ we are done.
  Assume that $l$ is odd. If $\gamma_{i,j}(c)= \lfloor l/2\rfloor$
  then $c_i=0$ and $c_j=0$. This yields $c_{i-1}=1$ and $c_{j+1}=1$
  otherwise Lemma~\ref{lem:basic_P1}.\ref{lem:basic_4} is violated.
  But in this case nodes $v_{i-1}$ to $v_{j+1}$ toggle their label in
  every round. This is impossible since $c\in \Cf$. Hence,
  $\gamma_{i,j}(c)\ge 1+ \lfloor l/2\rfloor=(l+1)/2 =\lceil
  l/2\rceil$.

  By assumption two neighboring nodes with threshold $2$ do not both
  have label $1$. Then $\gamma_{i,j}(c)\ge\lceil l/2\rceil$ implies
  that $c_i=1$ and $c_j=1$. If $c_{i-1}=0$ and $c_{j+1}=0$ then again
  the nodes $v_{i-1}$ to $v_{j+1}$ toggle their label in every round.
  This is impossible since $c$ converges. Thus,
  $\gamma_{i-1,j+1}(c) \ge \lceil l/2\rceil + 1$.

  It remains to consider the case that $l$ be even and $l\ge 4$. By
  our assumption we have $c_i=c_{i+2}=\ldots = 0$ and
  $c_{i+1}=c_{i+3}=\ldots = 1$, in particular $c_j=1$. If $c_{i-1}=0$
  then after one round $c_i=c_{i+1}=0$, contradicting
  Lemma~\ref{lem:basic_P1}.\ref{lem:basic_4}. If $c_{i-1}=1$ and
  $c_{j+1}=0$ then nodes $v_{i-1},\ldots,v_{j+1}$ keep toggling their
  labels. This is impossible since $c$ converges. Thus,
  $c_{i-1}=c_{j+1}=1$ and hence
  $\gamma_{i-1,j+1}(c) \ge \lceil l/2\rceil +2$.
    \end{enumerate}
  \end{proof}

\subsection{Proofs of Section~\ref{sec:case-r_fp=n_2}}

  \subsubsection{Proof of Lemma~\ref{lem:even_case}}
\label{sec:proof-lemma-even_case}
\begin{proof}
    The case $n=2$ is obvious. Let $n\ge 4$. First assume that
    $f\in {\cal E}_n$. By definition $f$ has no 1-streak of length
    $2$. Assume that $f$ has a 1-streak of length $0$ and index $i$
    saturated by some $c\in \Cfp{n/2}$. Then $i\equiv 0(2)$ by
    Lemma~\ref{lem:splittable} and Theorem~\ref{theo:lowerBound}.
    Since $f\in {\cal E}_n$ we have $f_j=2$ for all $j\le i$ or all
    $j\ge i$. Then Lemma~\ref{lem:basic_P1}.\ref{lem:basic_4} implies
    $r_f(P) > n/2$, a contradiction.

    Now assume that $f$ has no 1-streak of length $0$ saturated by a
    $c\in \Cfp{n/2}$ and no 1-streak of length $2$. We will prove that
    exactly one of the nodes $v_1,v_3,v_{n-2},v_n$ has threshold $1$,
    while all other $n-1$ nodes have threshold $2$. By
    Corollary~\ref{cor:cor0} $f$ has no 1-streak of length $l\ge 3$.
    Let $c\in \Cfp{n/2}$. The proof is split into three cases
    depending on the number of nodes with threshold $1$.

  \noindent
  \textbf{Case 1}: No node has threshold $1$, i.e., $f_i=2$ for all $i$.\\
  Then $n/2=\gamma(c)\ge \left\lceil (n+1)/2\right\rceil$ by
  Lemma~\ref{lem:basic_P1}.\ref{lem:basic_4}, a contradiction since $n$
  is even. Thus, this case cannot occur.\\[2mm]
  \textbf{Case} 2: There exists $k\ge 1$ such that all nodes except node $v_k$ have threshold $2$.\\
  Let $l=k-1$ and $t=n-k$. Then $l$ (resp.\ $t$) is the number of
  leading (resp.\ trailing) nodes with threshold $2$. First consider
  the case that $l$ is even. Then $t$ is odd. By
  Lemma~\ref{lem:basic_P1}.\ref{lem:basic_4} we have
  \[n/2=\gamma(c)\ge l/2 + c_k + (t+1)/2= c_k + n/2.\] This yields
  $c_k=0$ and $l/2$ of the leading nodes have label $0$. Thus, the
  leading $l$ nodes have labels $1,0,1,\ldots,0$. If $l\ge 4$ then after
  one round $c_{l-1}=c_{l}=0$, contradicting
  Lemma~\ref{lem:basic_P1}.\ref{lem:basic_4}. Thus, $l=2$ or $l=0$, i.e.,
  $f_3=1$ or $f_1=1$. If $l$ is odd, then $t$ is even. We repeat the
  above argument by replacing $l$ with $t$. This shows that either
  $f_n=1$ or $f_{n-2}=1$. Thus, $f=F^{i,n}$ with $i=1,3,n-2$, or $n$.\\[2mm]
  \textbf{Case} 3: There are at least two nodes with threshold $1$.\\
  We will prove that this case is impossible. Let $d$ be the minimum
  number of nodes between two nodes with threshold $1$. Let $v_i$ and
  $v_j$ be the first pair of such nodes. First, assume $d=0$ and let
  $l$ be the number of consecutive nodes including $v_i$ and $v_j$
  with threshold $1$, i.e., $l\ge 2$. Hence, there exists a 1-streak
  of $f$ of length $2$, a contradiction. Hence $d>0$. The $d+2$ nodes
  from $v_i$ to $v_j$ form a subsequence $S$ of type $A$.

  If $d$ is odd then $\gamma_{c_S}(c)\ge (d+1)/2+1 >\abs{S}/2$ by
  Lemma~\ref{lem:basic_P1}.\ref{lem:basic_5}. Now
  Lemma~\ref{lem:basic_case} implies $\gamma(c) > n/2$, a
  contradiction. If $d$ is even and $d\ge 4$ then
  $\gamma_{c_S}(c)\ge d/2+2 =\abs{S}/2 +1$ by
  Lemma~\ref{lem:basic_P1}.\ref{lem:basic_5}. Again
  Lemma~\ref{lem:basic_case} implies $\gamma(c) > n/2$, a
  contradiction.

  Thus, only the case $d=2$ remains. Then $f_i=f_{i+3}=1$ and
  $f_{i+1}=f_{i+2}=2$ for some $i$. We claim that this is also
  impossible. Since $f$ has no saturated 1-streak for $c$ of length
  $0$ we have $c_{i+1}\not=c_{i+2}$, hence $c_{i+1}=1$ or $c_{i+2}=1$.
  By Lemma~\ref{lem:basic_P1}.\ref{lem:basic_5} and
  Lemma~\ref{lem:basic_case} we have $\gamma_{c_S}(c)= 2$ and thus,
  $c_{i}=1$ or $c_{i+3}=1$.

  We assume that $c_i=1$ (the case $c_{i+3}=1$ is proved in the same
  way). If $c_{i+1}=0$ then $c_{i+2}=1$ and $c_{i+3}=0$. This is
  impossible since $c\in \Cf$. Thus, we have $c_{i+1}=1$. Then
  $c_{i+2}=c_{i+3}=0$. This requires $i+3<n$. Since $d>0$ we have
  $f_{i+4}=2$ and $c_{i+4}=1$. First consider the case that $i>1$.
  Then clearly, $f_{i-1}=2$. Assume that $c_{i-1}=1$. Thus, $f$ has a
  1-streak of length $1$ and index $i$ that is saturated by
  $c$, a contradiction.
  Thus, $c_{i-1}=0$. Then after one round
  the nodes corresponding to $S$ have label $0,1,0,1$. This is
  impossible since $c\in \Cf$. It remains
  to consider the case $i=1$. Remember $c_1=c_2=1$. Clearly,
  $c|_{\vec{P}_2}$ converges for $f$ to $\mathbb{1}$. Thus, we have
  \[n/2=\gamma(c) = 1 + \gamma_{c_{\vec{P}_2}}(c)\ge 1+ n/2\] by
  Theorem~\ref{theo:lowerBound}, a contradiction. 
\end{proof}

\subsubsection{Proof of Theorem~\ref{theo:even_case_2}}
\label{sec:proof-theo:even_case_2}

\begin{proof}
  First let $f\in {\cal E}_n^\ast$. If $f\in {\cal E}_n$ then by
  Lemma~\ref{lem:even_case} no 1-streak of $f$ of length $0$ is
  saturated by a $c\in \Cfp{n/2}$. If $f=F^{n/2,n/2}\circ F^{1,n/2}$
  then the argument used in the first part of the proof of
  Lemma~\ref{lem:even_case} yields again that no 1-streak of $f$ of
  length $0$ is saturated by a $c\in \Cfp{n/2}$.

  Next, assume that $f$ has no 1-streak of length $0$ saturated by a
  $c\in \Cfp{n/2}$. If $f$ has no 1-streak of length $2$ then the
  result follows from Lemma~\ref{lem:even_case}. Assume that $f$ has
  1-streak of length $2$ and index $i$. Then $n\ge 4$ and
  \[r_f(P) \ge r_{f|_{\cev{P}_{i}}}(\cev{P}_{i}) +
    r_{f|_{\vec{P}_{i+1}}}(\vec{P}_{i+1}) \ge \left\lceil
      i/2\right\rceil + \left\lceil (n-i)/2\right\rceil\] by
  Lemma~\ref{lem:splittable_2} and Theorem~\ref{theo:lowerBound}.
  Assume that $i$ is odd. Then $n/2\ge (n+2)/2$, a contradiction.
  Hence, $i$ is even. Assume that $f$ has another 1-streak of length
  $2$ with index $i'\ge i+3$. We choose the index with the smallest
  possible $i'$. Using the same argument proves that $i'$ is even. Let
  $P'=P_{i+1,i'}$. Clearly $\abs{P'}$ is even and $f|_{P'}$ has no
  saturated 1-streak by $c|_{P'}$ for any length $l \ge 0$. The first
  and the last node of $P'$ have threshold $1$. But by
  Lemma~\ref{lem:even_case} there is no such threshold function.
  Hence, there exists just one 1-streak of length $2$ saturated by
  $c$.

  Hence, both paths $\cev{P}_{i}$ and $\vec{P}_{i+1}$ have even length
  and $c|_{\cev{P}_{i}}$ (resp. $c|_{\vec{P}_{i+1}}$) converge to
  $\mathbb{1}$ for $f|_{\cev{P}_{i}}$ (resp.\ $f|_{\vec{P}_{i+1}}$) by
  Lemma~\ref{lem:splittable_2} and satisfy the assumption of
  Lemma~\ref{lem:even_case}. Thus, we have
  $f|_{\cev{P}_{i}}=(2,\ldots,2,1)$ and
  $f|_{\vec{P}_{i+1}}=(1,2,\ldots,2)$ and the corresponding labelings
  $c|_{\cev{P}_{i}}$ and $c|_{\vec{P}_{i+1}}$ converge in $i-1$ resp.\
  $n-i-1$ rounds to $\mathbb{1}$. If $i-1< n-i-1$ then the labeling
  for $\cev{P}_{i}$ converges before the labeling for $\vec{P}_{i+1}$.
  This has the effect that after round $i-1$ nodes $v_{i+1}$ and
  $v_{i+2}$ toggle their label in every round. This is impossible
  since $c$ converges. This yields $i-1=n-i-1$ and $i=n/2$. Since $i$
  is even we have $n\equiv 0(4)$ and all but the nodes $v_{n/2}$ and
  $v_{n/2+1}$ have threshold $2$.
\end{proof}

\subsubsection{Lemma~\ref{lem:bound2_1}}
\label{sec:proof-lemma-refl}

\begin{lemma}\label{lem:bound2_1} Let $P$
  be a path with an odd number $n$ of nodes, $(P,f)$ an $f$-reversible
  process, and $c\in \Cf(\left\lceil n/2\right\rceil)$. If $f$ has no
  saturated 1-streak for $c$ of length $0$ or $1$ and no 1-streak of
  length $2$ then there are at most two nodes with threshold
  $1$. 
  \end{lemma}
\begin{proof}
  Let $v_{i_1},v_{i_2},\ldots, v_{i_k}$ be the nodes with threshold
  $1$. Then $i_j +1 < i_{j+1}$ by assumption. Assume that $k\ge 4$.
  Let $P_j=P_{i_j,i_{j+1}}$. First assume that $\abs{P_{1}}$ is odd.
  Then $\gamma(c_{P_{1}}) = (\abs{P_1}+1)/2$ by
  Lemma~\ref{lem:basic_P1}.\ref{lem:basic_5}. If $\abs{P_j}$ is for
  some $j>1$ also odd then
  $\gamma(c_{i_j+1,i_{j+1}-1}) = (\abs{P_j}-1)/2$ by the same Lemma.
  Then by Lemma~\ref{lem:basic_case} $\gamma(c) > (n+1)/2$, a
  contradiction. Hence, $\abs{P_j}=4$ for $j>1$ by
  Lemma~\ref{lem:basic_P1}.\ref{lem:basic_5}. Consider the path
  $v_{i_2}, \ldots, v_{i_4}$ of length $7$. The corresponding
  threshold values are $1,2,2,1,2,2,1$. By
  Lemma~\ref{lem:basic_P1}.\ref{lem:basic_7} the only labelings for
  nodes $v_{i_2}, \ldots, v_{i_3}$ and $v_{i_3}, \ldots, v_{i_4}$ are
  $0,0,1,1$ and $1,1,0,0$. Thus, the only two possible labelings for
  $v_{i_2}, \ldots, v_{i_4}$ are $0,0,1,1,1,0,0$ and $1,1,0,0,0,1,1$.
  The first is impossible because in this case there is a 1-streak of
  length $1$ that is saturated by $c$ and the second because
  $c\in \Cf$. Hence, $k\le 3$. Suppose $k=3$. Then $P_{i_1,i_3}$ has
  even number of nodes. Since $n$ is odd, $f$ has an odd number of
  trailing or leading $2$'s. With
  Lemma~\ref{lem:basic_P1}.\ref{lem:basic_4} this leads again to a
  contradiction to Lemma~\ref{lem:basic_case} if the number of leading
  $2$'s is odd. If this number is even then it is easy to see that $c$
  does not converge. This contradiction shows that $k\le 2$.

  Next, assume that $\abs{P_{1}}$ is even. By
  Lemma~\ref{lem:basic_P1}.\ref{lem:basic_5} we have $\abs{P_1}=4$. By
  the above argument it is impossible that $\abs{P_{2}}$ is also even.
  Thus, $\abs{P_{2}}$ is odd and $\abs{P_{j}}=4$ for $j\ge 3$. As
  above we infer that $k\le 4$. Assume $k=4$. Since
  $\gamma(c_{P_{2}}) = (\abs{P_2}+1)/2$ by
  Lemma~\ref{lem:basic_P1}.\ref{lem:basic_5} the number of trailing
  and leading $2$'s of $f$ must be even (since $n$ is odd). Hence,
  $c_{i_1-1}=c_{i_4+1}=0$. Lemma~\ref{lem:basic_P1}.\ref{lem:basic_7}
  implies that $c_{i_2}=c_{i_3}=1$ and hence,
  $c|_{P_2}=(1,0,1, \ldots, 0,1)$. This, labeling does not converge,
  a contradiction. Hence, $k\le 3$. As in the first case this implies
  $k\le 2$.
\end{proof}

\subsubsection{Proof of Lemma~\ref{theo:odd_case}}
\label{sec:proof-theo:odd_case}

\begin{proof}
  We consider two cases. Let
  $c\in \Cf(\left\lceil n/2\right\rceil)$.
  
  \textbf{Case 1}: $f$ has no 1-streak of length $2$. By
  Lemma~\ref{lem:bound2_1} (see Appendix~\ref{sec:proof-lemma-refl})
  at most two nodes of $P$ have threshold $1$. If at most one node has
  threshold $1$ then $f_i=2$ for all nodes or $f_i=2$ for all nodes
  but one. Hence, $f=F^{k,n}$ for $k=1,\ldots,n$ or $f=F^n$, i.e.,
  $f\in {\cal O}_n^\ast$. Thus, we can assume that $f$ is of the form
  $B_1 \circ A \circ B_2$, with $f_i=f_j=1$.

    First consider the case that $\abs{A}$ is odd. By
    Lemma~\ref{lem:basic_P1}.\ref{lem:basic_5} we know that
    $\gamma(c_A)=(\abs{A}+1)/2$. Then $\abs{B_1}$ and $\abs{B_2}$ are
    either both odd or both even. The first case is impossible, since
    Lemma~\ref{lem:basic_P1}.\ref{lem:basic_4} would imply
  \[\frac{n+1}{2}=\gamma(c)=\gamma(c_{B_1}) + \gamma(c_{A}) +
    \gamma(c_{B_2})\ge \frac{n+3}{2}.\] Hence, $\abs{B_k}\equiv 0(2)$
  for $k=1,2$. Thus, $\gamma(c_{B_k})=\abs{B_k}/2$ by
  Lemma~\ref{lem:basic_P1}.\ref{lem:basic_4}. If $\abs{B_1}>0$ then
  $c_1=1$. Assume that $\abs{B_1}\ge 4$. Then $c_{i-3}=c_{i-1}=0$ and
  $c_{i-2}=1$. If $c_i=0$ then after one round $c_{i-2}=c_{i-1}=0$,
  this contradicts Lemma~\ref{lem:basic_P1}.\ref{lem:basic_4}. Hence
  $c_i=1$. Since $\gamma(c_{A})=(\abs{A}+1)/2$ we have $c_j=0$,
  otherwise $\gamma(c) > (n+1)/2$. This argument also shows that
  $\abs{B_2}<4$, i.e., $\abs{B_2}$ either equals $0$ or $2$. In the
  first case this yields $j=n$ and $i=5,7,\ldots,n-2$, i.e.,
  $F^{k,k} \circ F^{n-k,n-k}$ and in the latter case this leads to
  $j=n-2$ and $i=5,7,\ldots,n-4$, i.e., $F^{k,k} \circ F^{n-k-2,n-k}$.
  It remains the case that $\abs{B_1}$ is $0$ or $2$. Then $i=1$ and
  $j=3,5,\ldots,n$, i.e., $F^{1,k}\circ F^{1,n-k}$ for
  $k=2,4,\ldots,n-1$ or $i=3$ and $j=5,7,\ldots,n-2$, i.e.,
  $F^{3,k} \circ F^{1,n-k}$ for $k=4,6,\ldots,n-1$.

  Finally, we consider the case that $\abs{A}$ is even. Then either
  $\abs{B_1}\equiv 1(2)$ or $\abs{B_2}\equiv 1(2)$. By
  Lemma~\ref{lem:basic_P1}.\ref{lem:basic_5} we have $A=(1,2,2,1)$ .
  First assume that $\abs{B_1}\equiv 1(2)$ and $\abs{B_2}\equiv 0(2)$.
  If $\abs{B_2}>0$ then the labeling of $A$ is by
  Lemma~\ref{lem:basic_P1}.\ref{lem:basic_7} either $(0,0,1,1)$ or
  $(1,1,0,0)$, note that $f$ has no saturated 1-streak of length $0$.
  Hence the labeling of $2\circ A \circ 2$ is either $(1,0,0,1,1,0)$
  or $(1,1,1,0,0,0)$. This is impossible since in both cases it would
  imply $c\not\in C_f$. Hence, $\abs{B_2}=0$. Thus, $f_{n-3}=f_n=1$,
  i.e., $F^{n-3,n-3} \circ F^{3,3}$. Similarly if
  $\abs{B_1}\equiv 0(2)$ and $\abs{B_2}\equiv 1(2)$ then
  $f_{1}=f_4=1$, i.e., $F^{1,3}\circ F^{1,n-3}$.

  \textbf{Case 2}:  $f$ has a 1-streak of length $2$. Let $i$ be the
  smallest index of a 1-streak of length $2$. Note that it cannot be
  saturated by $c$. Thus, $c_i=c_{i+1}=0$. By
  Lemma~\ref{lem:splittable_2} and Theorem~\ref{theo:lowerBound} we
  have
  \[(n+1)/2=\gamma(c) = \gamma(c|_{\cev{P}_{i}}) +
  \gamma(c|_{\vec{P}_{i+1}}) \ge \left\lceil i/2\right\rceil +
  \left\lceil (n-i)/2\right\rceil = (n+1)/2.\] Hence,
  $r_{f|_{\cev{P}_{i}}}(\cev{P}_{i})=\left\lceil i/2\right\rceil$ and
  $r_{f|_{\vec{P}_{i+1}}}(\vec{P}_{i+1})=\left\lceil
    (i+1)/2\right\rceil$. Assume that $i$ is even. Then
  $f_{\cev{P}_{i}}=(2,\ldots,2,1)$ by Theorem~\ref{theo:even_case_2}.
  Assume that there exists another 1-streak of length $2$. Let $j>i$
  be the smallest index of such a 1-streak. Then again $c_j=c_{j+1}=0$
  and Theorem~\ref{theo:even_case_2} applied to $P_{i+1,j}$ implies
  that $j$ is odd, because if $j$ would be even then $c_{i+1}$ and
  $c_{j}$ cannot be $0$ at the same time. Thus, $P_{i+1,j}$ satisfies
  the assumptions of Lemma~\ref{theo:odd_case}, hence $c_j=1$.
  This contradiction implies that there exists no second 1-streak of
  length $2$. Then Theorem~\ref{theo:even_case_2} yields a description
  of $f_{\cev{P}_{i}}$ and Lemma~\ref{theo:odd_case} a description
  of $f_{\vec{P}_{i+1}}$. Note that some combinations of
  $f_{\cev{P}_{i}}$ and $f_{\vec{P}_{i+1}}$ according to
  Theorem~\ref{theo:even_case_2} and Lemma~\ref{theo:odd_case} do not
  converge. The reason is that it can lead to a configuration, in which
  $c_i\not=c_{i+1}$ holds, a contradiction. It is easy to verify that
  only combinations lead to convergence to $\mathbb{1}$ in which the
  three nodes $v_{2i},v_{2i+1}$ and $v_{4i+1}$ have threshold $1$,
  here $i=1,\ldots, \left\lfloor (n-1)/4\right\rfloor$. If $i$ is odd,
  we can repeat the above argument leading to threshold functions that
  are inverse to the described functions.
\end{proof}

\end{document}